\documentclass[pra,superscriptaddress,nofootinbib,noshowpacs,preprintnumbers,longbibliography,floatfix]{revtex4-2}
\usepackage[utf8]{inputenc}
\usepackage{graphicx}
\usepackage{float}
\usepackage{amssymb}
\usepackage{amsmath}  
\usepackage{adjustbox}
\usepackage{mathtools}
\usepackage{dsfont}
\usepackage{array}
\usepackage{bm,fixmath}
\usepackage{mathrsfs}
\usepackage{pifont}
\usepackage{multirow}
\usepackage{upgreek}
\usepackage{caption}
\usepackage{subcaption}
\usepackage{xcolor}
\usepackage{bm}
\usepackage{bbm}
\usepackage{physics}
\usepackage{comment}
\usepackage{tikz}
\usetikzlibrary{quantikz}
\usepackage[paperwidth=240mm,paperheight=297mm,centering,hmargin=3cm,vmargin=3cm]{geometry}

\usepackage[colorlinks = true,
            linkcolor = blue,
            urlcolor  = blue,
            citecolor = violet,
            anchorcolor = blue]{hyperref}

\usepackage{orcidlink} 

\begin{document}

\title{Quantum computing inspired paintings: reinterpreting classical masterpieces}

\author{Arianna~Crippa \orcidlink{0000-0003-2376-5682}}
\email{cripparianna@gmail.com}
\affiliation{Deutsches Elektronen-Synchrotron DESY, Platanenallee 6, 15738 Zeuthen, Germany
}
\affiliation{Institut für Physik, Humboldt-Universität zu Berlin, Newtonstr. 15, 12489 Berlin, Germany}

\author{Yahui~Chai \orcidlink{0000-0002-3404-6096}}
\affiliation{Deutsches Elektronen-Synchrotron DESY, Platanenallee 6, 15738 Zeuthen, Germany
}

\author{Omar~Costa~Hamido~\orcidlink{0000-0001-5077-853X}}
\affiliation{Centre for Interdisciplinary Studies (CEIS20), University of Coimbra, Portugal}

\author{Paulo~Itaborai \orcidlink{0000-0002-4956-2958}}
\affiliation{
 Computation-Based Science and Technology Research Center, The Cyprus Institute, 20 Kavafi Street,
2121 Nicosia, Cyprus
}
\affiliation{Deutsches Elektronen-Synchrotron DESY, Platanenallee 6, 15738 Zeuthen, Germany
}

\author{Karl~Jansen~\orcidlink{0000-0002-1574-7591}}
\email{karl.jansen@desy.de}
\affiliation{
 Computation-Based Science and Technology Research Center, The Cyprus Institute, 20 Kavafi Street,
2121 Nicosia, Cyprus
}
\affiliation{Deutsches Elektronen-Synchrotron DESY, Platanenallee 6, 15738 Zeuthen, Germany
}

\collaboration{QC-PAINT}
\date{\today}

\begin{abstract}

We aim to apply a quantum computing technique to compose artworks. The main idea is to revisit three paintings of different styles and historical periods: ``\textit{Narciso}'', painted circa 1597–1599 by Michelangelo Merisi (Caravaggio), ``\textit{Les fils de l'homme}'', painted in 1964 by René Magritte and ``\textit{192 Farben}'', painted in 1966 by Gerhard Richter. 
We utilize the output of a quantum computation to change the composition in the paintings, leading to a paintings series titled ``\textit{Quantum Transformation I, II}, \textit{III}''.
In particular, the figures are discretized into square lattices and the order of the pieces is changed according to the result of the quantum simulation. We consider an Ising Hamiltonian as the observable in the quantum computation and its time evolution as the final outcome.
From a classical subject to abstract forms, we seek to combine classical and quantum aesthetics through these three art pieces.
Besides experimenting with hardware runs and circuit noise, our goal is to reproduce these works as physical oil paintings on wooden panels. With this process, we complete a full circle between classical and quantum techniques and contribute to rethinking Art practice in the era of quantum computing technologies. 
\end{abstract}

\maketitle

\section{Introduction}
Art and science have long gone hand in hand, and quantum mechanics has attracted the interest of artists from the very beginning. However, even though quantum computing technologies have slowly started to emerge in the late 90s~\cite{QuantumComputing2000}, their use in the Arts has only started sprouting recently. This integration has been proposed in the visual arts~\cite{lioretQuantumArt2016} and in music~\cite{mirandaQuantumComputerMusic2022a}, and has inspired several collaborations, see e.g. Ref.~\cite{Clemente:2022qce,Itaborai:2023xha,Itaborai:2024fxj} and exhibitions \cite{qiskitMakingInvisibleVisible2021,Quantumblur}. The interplay between being inspired by quantum and making use of quantum computing has been discussed in Ref.~\cite{AdventuresQuantumland2021}, where it was also proposed the concept of Quantum-Computing Aided Composition (QAC) and Quantum-Computing Aided Design (QAD) to articulate the use of quantum technologies in Art. In our work we add another step to this approach by using the quantum computer-generated image as a template to actually paint the digital quantum computer result on a real canvas. In this way, a human element is added to the process, leading also to wanted inaccuracies, reflecting the human part of the process. 
Here, we present a detailed example of this idea to use quantum computing to aid the creative process of human artistic oil painting.
In particular, in this project, we consider three masterpieces: ``\textit{Narciso}'', painted circa 1597–1599 by Michelangelo Merisi (Caravaggio)~\cite{narciso}, ``\textit{Les fils de l'homme}'', painted in 1964 by René Magritte~\cite{magritte} and ``\textit{192 Farben}'', painted in 1966 by Gerhard Richter~\cite{192colors}.
We use quantum computing, in particular the IBM quantum hardware~\cite{ibmq}, to process digital images. To develop the quantum algorithm we use IBM's quantum computing programming framework, \textit{Qiskit}~\cite{qiskit2024}.
The original paintings are analyzed and divided partly or as a whole into a lattice. The discrete units, i.e. the lattice tiles, are then encoded to qubits, the fundamental constituents of a quantum computer. We then employ a Hamiltonian of a physical system as the observable we measure: the Ising model~\cite{ising1924post}. By quantum time evolving this operator, we can follow the transformation of the initial lattice 
from a selected starting row or column of the lattice. Running the algorithm on a real quantum computer results in a quantum ordering of the lattice tiles. This means that the tiles are placed only with a certain probability given by the underlying quantum mechanical principles of superposition and entanglement. 
In this way, the obtained digital image generated by the quantum computer serves, in a twofold way, as a template for the human interface. First, from a number of digital examples the one that finds the consent of the authors as aesthetically the best is selected as the final template. Second,   
the selected digital image is then taken to be reproduced manually on wooden panels, using oil paint. As said above, in this way we are realizing our idea of keeping the artwork still ``human''.

Our choice of paintings follows a line from the Naturalism of Caravaggio to the Surrealism of Magritte and, lastly, the Abstraction of Richter. We consider our work from various angles. The first is to explore, how quantum computing renders the original paintings from the three epochs differently leading to new versions of the paintings. The second is the aesthetical reception of the digital images produced by the authors. And, the third is the human intervention by reproducing the digital template on a wooden panel with oil, giving it an additional human element. Finally, through using a quantum computer, and thus employing the quantum mechanical principles, we are able to represent a contemporary concept, i.e. how reality is losing its form and becoming abstract and evanescent. 
We titled the paintings series as ``\textit{Quantum Transformation I, II} and \textit{III}''\footnote{The paintings will be exhibited at the Ars Electronica Festival 2025~\cite{ars_electr}.}.

\vspace{0.2cm}
The paper is structured as follows: In Sec.~\ref{sec:methods}, we introduce the Hamiltonian of the Ising model and the time evolution via a Trotterization. We define the quantum circuit used for the computation with its parameterized gates (i.e. the building blocks of quantum circuits). The results are reported in Sec.~\ref{sec:narciso}, for the painting ``\textit{Quantum Transformation I: Caravaggio}'', Sec.~\ref{sec:magritte} for the painting ``\textit{Quantum Transformation II: Magritte}'', and Sec.~\ref{sec:richter} for the painting ``\textit{Quantum Transformation III: Richter}''. In each section, we describe each of the three paintings, with the results from the quantum computation, and show the final pictures of the oil paintings. The source code used to work on these paintings is available in Ref.~\cite{qcpaint_sourcecode}. To conclude the paper, Sec.~\ref{sec:conclusion} gives a summary and the outlook for this project.

\section{Numerical methods}\label{sec:methods}
The Ising model is a particular example of an interesting lattice spin system, which exhibits a quantum phase transition. Introduced as a model of ferromagnetism~\cite{ising1924post}, it has also been used to study collective phenomena in various non-physical systems, e.g. a work by part of the authors on the flight gate assignment problem on a trapped ion quantum computer~\cite{chai2023simulatingflightgateassignment}.
In this paper we utilize this model, using a time evolution, to modify the above-listed paintings in order to produce new artworks.
In the quantum computing analysis, we consider the Ising Hamiltonian,
\begin{equation}\label{eq:ising_H}
    H = \sum_{n=0}^{N-1} J_n Z_n Z_{n+1} + \sum_{n=0}^{N-1} h_{z,n} Z_n +  \sum_{n=0}^{N-1} h_{x,n}  X_n
\end{equation}
for a discrete system of $N$ spins, assuming values $\pm 1$, with periodic boundary conditions. In Eq.~\eqref{eq:ising_H} the parameters $J_n$ define the interaction between neighboring spins and $h_{z,n}$ or $h_{x,n}$ refer to external longitudinal and transversal magnetic fields, respectively. The operators $Z,X$ correspond to the Pauli matrices,
\begin{equation}
    Z=\begin{pmatrix}
        1 & 0\\
        0 & -1
    \end{pmatrix}, \ \ \ X=\begin{pmatrix}
        0 & 1 \\
        1 & 0
    \end{pmatrix}.
\end{equation}

We then use the Hamiltonian to calculate how the given quantum system evolves in time. Starting from an initial state at time $t=0$, $\ket{\psi(0)}$, the time evolution of this state, after a certain time $t$, can be written as the action of a unitary time evolution operator $U(t)$,
\begin{equation}
    \ket{\psi(t)}=U(t) \ket{\psi(0)} \equiv e^{-iHt} \ket{\psi(0)}.
\end{equation}

\begin{figure}[htp!]
    \centering
    \includegraphics[width=0.7\columnwidth]{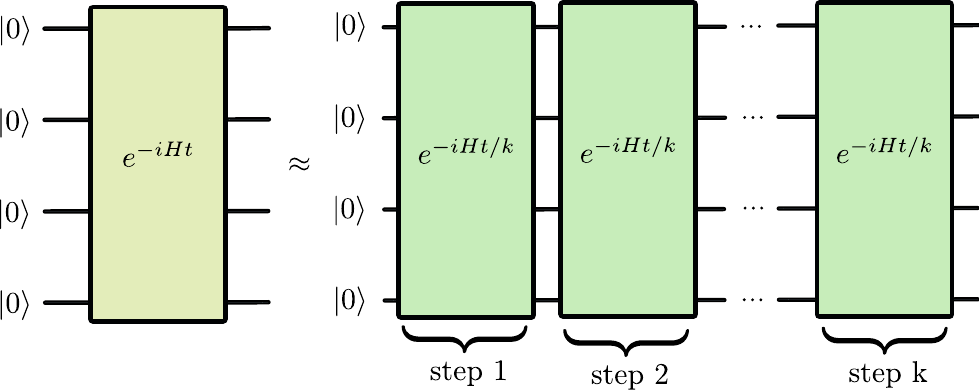}
    \caption{\textbf{Parameterized quantum circuit for the time evolution:} Example of $N=4$ qubits for the time evolution, computed with the Trotterization method. Each (time)step in the circuit is repeated $k$ times to reach the approximation in Eq.~\eqref{eq:u_t_approx}. Details on the gates on this quantum circuit are in Fig.~\ref{fig:trotter_circuit} and Fig.~\ref{fig:trotter_gates}.}
    \label{fig:trotter_timeevol_tot}
\end{figure}

\begin{figure*}[htp!]
    \centering
    \includegraphics[width=0.85\textwidth]{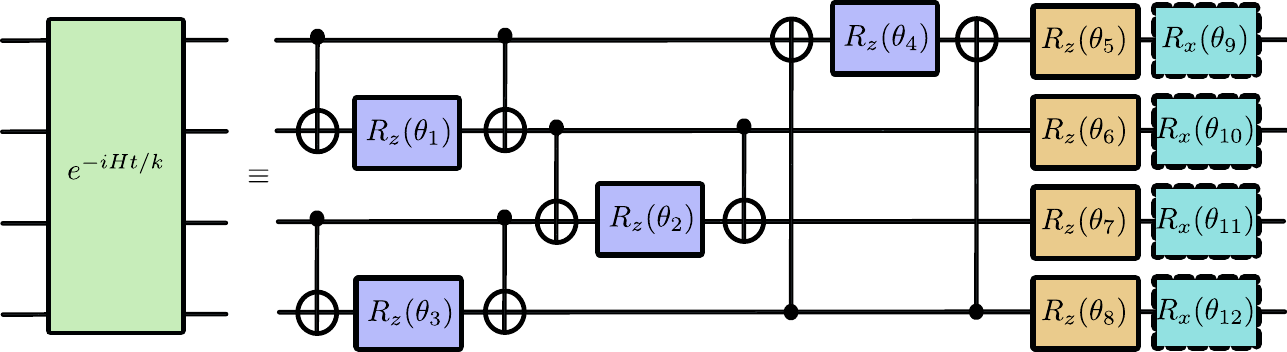}
    \caption{\textbf{Trotter step quantum circuit:} The steps in the time evolution of Fig.~\ref{fig:trotter_timeevol_tot} are represented by this circuit. The set of gates (example with $N=4$ qubits) refer to the $N$ terms in Eq.~\eqref{eq:ising_H}. Details of the gates in Fig.~\ref{fig:trotter_gates}.}
    \label{fig:trotter_circuit}
\end{figure*}

\begin{figure}[H]
    \centering
    \includegraphics[width=0.65\columnwidth]{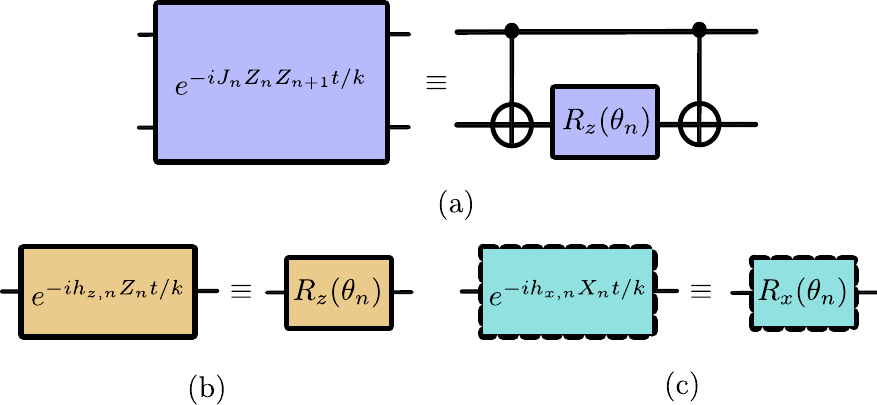}
    \caption{\textbf{Set of quantum gates for the time evolution:} The terms in the evolution of the Hamiltonian of Eq.~\eqref{eq:ising_H} are translated into rotational gates onto the quantum circuit. (panel (a)) $R_{zz}(\theta_n)$ related to $J_n Z_n Z_{n+1}$, (panel (b)) the $R_z(\theta_n)$ gate related to $h_{z,n}Z_n$ and (panel (c)) the $R_x(\theta_n)$ gate representing the term $h_{x,n}X_n$ of the Hamiltonian.}
    \label{fig:trotter_gates}
\end{figure}

Since a quantum circuit is a discrete system, we can only approximate the time evolution operator, and we employ a method called \textit{Trotterization}~\cite{suzuki1976generalized,trotter1959product}. For a generic Hamiltonian, that can be expressed as a sum of hermitian terms, such as Eq.~\eqref{eq:ising_H}, we can approximate $U(t)$ as
\begin{equation}
\label{eq:u_t_approx}
    U(t) \approx \prod^k \prod_{n=0}^{N-1}  e^{-iH_n t/k}
\end{equation}
with $\Delta t=t/k$ being the \textit{time step}.
We now show the circuit that performs $k$ time evolution steps in a quantum device. 
Here, and in the rest of the paper, we follow the \textit{right-left} ($\ket{..q_2q_1q_0}$), \textit{top-bottom} ordering of the qubits.
Firstly, we assign each spin of the Ising system to individual qubits, which are initialized with a $\ket{0}$ state. Then, the circuit depicted in Fig.~\ref{fig:trotter_timeevol_tot} (example with $N=4$ qubits) is applied, to increase by $k$ time steps the evolution (Eq.~\eqref{eq:u_t_approx}). Each step in the circuit is reproduced by Fig.~\ref{fig:trotter_circuit}. 
This circuit structure involves three types of parameterized gates: $R_{zz}(\theta_n)$, $R_z(\theta_n)$ and $R_x(\theta_n)$, depicted in panels (a), (b) and (c) of Fig.~\ref{fig:trotter_gates}, respectively. They represent the terms in the Hamiltonian in the exponent of the time evolution operator. In general we have $R_P(\theta_n)\equiv e^{i\frac{\theta_n}{2}P}$ with $P=X,Y,Z$ Pauli matrix, thus in this work $\theta_n$ will be translated into the prefactors in the time operator,
\begin{equation}
    R_{zz}(\theta_n)\equiv e^{-iJ_n Z_N Z_{n+1}t/k}, \ \ \ R_{z}(\theta_n)\equiv e^{-ih_{z,n} Z_N t/k},\ \ \ R_{x}(\theta_n)\equiv e^{-ih_{x,n} X_N t/k}.
\end{equation}

For the numerical analysis, we perform the time evolution on selected IBM quantum superconducting devices~\cite{ibmq}. These devices belong to the so-called Noisy Intermediate-Scale Quantum (NISQ) computers~\cite{preskill2018quantum}, i.e. quantum devices that operate at a scale of tens to a few hundred qubits and are subject to quantum noise and hardware errors.
In this project, we have decided to work on the raw output data from the quantum computation, by not applying any error correction or mitigation techniques.
The idea is to create a lattice of the painting under study and get a new order of the lattice tiles from the unmitigated quantum results.

\section{Quantum Transformation I: Caravaggio}\label{sec:narciso}
In this section we describe the result of the first masterpiece we have considered: ``\textit{Narciso}''~\cite{narciso}, by the Italian artist Michelangelo Merisi, known as ``Caravaggio''.
The original composition, painted circa 1597–1599, has, as its only protagonist, an adolescent boy dressed in an elegant brocade doublet and leaning on both hands over the very dark surface of a body of water as he contemplates his distorted reflection~\cite{poseq1991allegorical}.
The main theme is Ovid's story of Narcissus, who, unable to tear himself from his image reflected, died consumed by his passion.

\begin{figure}[ht!]
    \centering
    \includegraphics[width=1\linewidth]{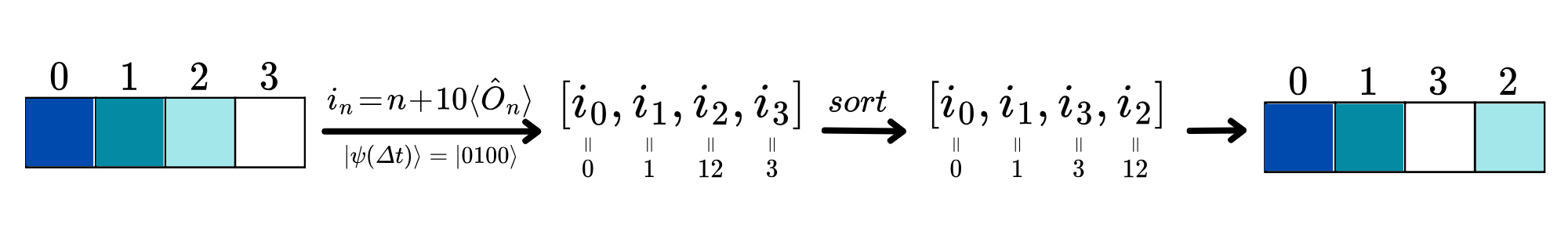}
    \caption{\textbf{Depiction of the reordering based on the quantum observables:} Example of the reordering of the numerical model for $t=\Delta t$.}
    \label{fig:example}
\end{figure}

\begin{figure}[H]
    \centering
    \includegraphics[width=0.7\columnwidth]{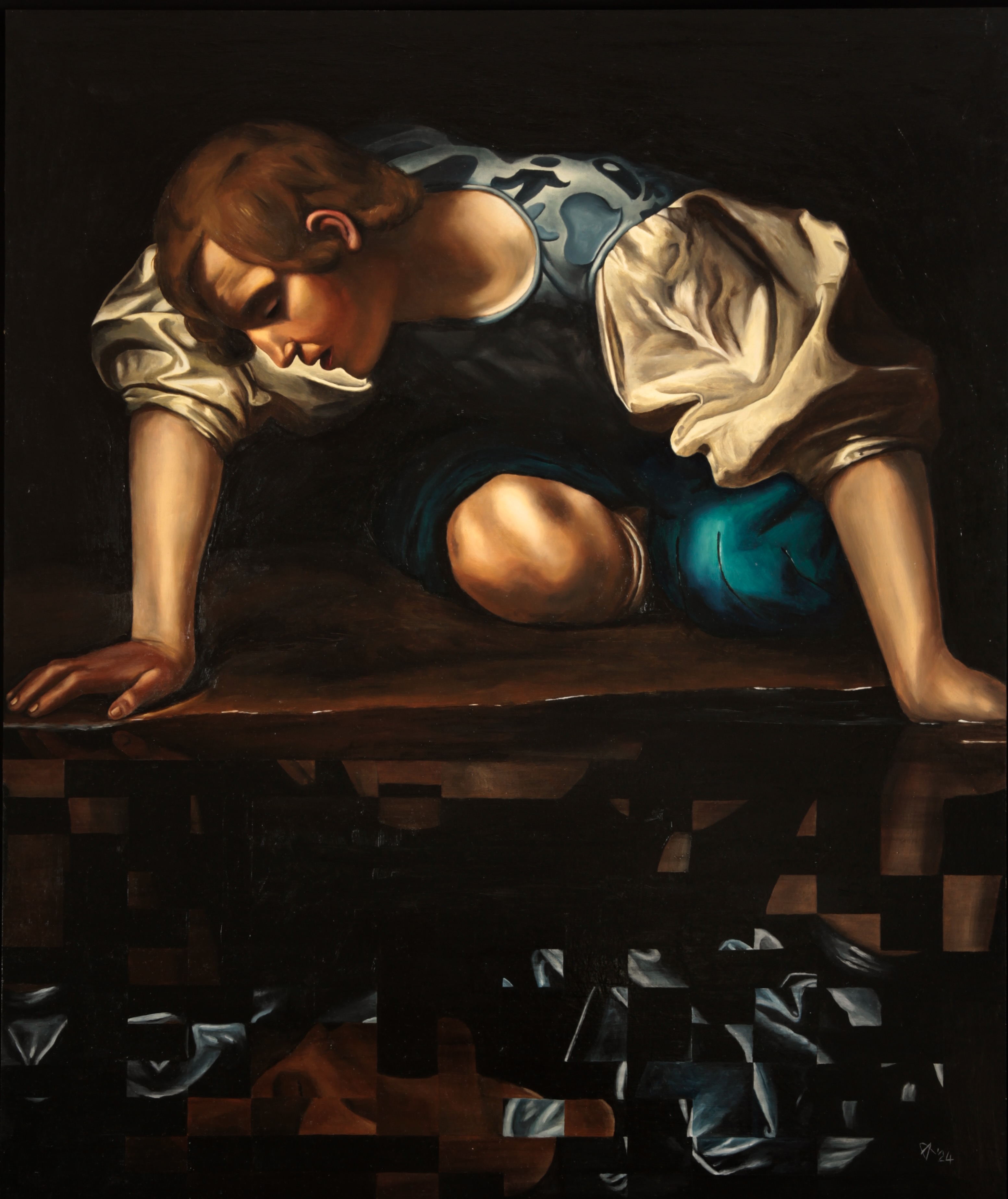}
    \caption{\textbf{The painting ``\textit{Quantum Transformation I: Caravaggio}'' (oil on wooden panel):} This photograph of the oil painting illustrates how the reflection (\textit{lower part}) has been modified by translating the results from quantum computation. \textit{Panel size: $70\times 84$ cm.}  }
    \label{fig:narcissus}
\end{figure}

In our revisited version, depicted in Fig.~\ref{fig:narcissus}, we still see a young boy looking intensely at his reflection. 
However, he is now observing a new chaotic shape, changed by the results of a quantum computation. 
This newly disturbed water also engages in analogy to the wave function collapse, where the superposition of multiple eigenstates, the different possible paths for the time evolution, get reduced, by means of observation -- as the boy does -- to what we can actually observe and measure. 
To compose this painting, we divide the lower part of the composition, depicting Narcissus' reflection, into a grid of 16 columns and 13 rows. 
By considering a time evolution based on the Hamiltonian in Eq.~\eqref{eq:ising_H}, with random couplings ($J_n,h_{z,n},h_{x,n}$), we proceed through steps. A row with $N$ squares will need the same number of qubits to be simulated. 
We measure the expectation value of the Pauli $Z$ for each qubit and use the results to define the new order of the elements in the painting.
Each column represents a qubit and the rows are the steps in the time evolution. 
 The initial state is $\ket{\psi(0)} = \ket{0}^{\otimes N}$ (with $N=16$), and we use observables defined as 
\begin{equation}\label{eq:obs_On}
    \hat{O}_n = \frac{(I-\sigma_n^z)}{2},
\end{equation}
with $n \in [0,N-1]$. For each time slice, we compute and order the value 
\begin{equation}
    i_n = \left( n + 10 \cdot \bra{\psi(t)} \hat{O}_n \ket{\psi(t)} \right)
    \label{eq:reorder}
\end{equation}
for all sites. 
In turn, the list produced, \{\(i_0\), \(i_1\), ... \(i_{N-1}\)\}, is sorted in ascending order, which determines the reordering of elements in the corresponding painting. 
The action of this operator on a $n$th qubit state is
\begin{align}
    \hat{O}_n \ket{0}=0 \ \ \ \text{and} \ \ \ \hat{O}_n \ket{1}=\ket{1}.
\end{align}
In particular, 
\begin{equation}
    \bra{\psi(0)} \hat{O}_n \ket{\psi(0)} = 0 \ \ \ \text{at} \ \ t=0,
    \label{eq:zero}
\end{equation} 
for all sites, so the order of the elements in the painting remains unchanged in this initial time slice. 
In summary, the image is first indexed. Then, Eq.~\eqref{eq:reorder} produces a list of values \(i_n\) which, in turn, when sorted, determines the new positions of the elements.
The factor `\(10\)' in Eq.~\eqref{eq:reorder} is introduced so that the order of magnitude of the computed expectation value is comparable with the integer indexes \(n\). In simple terms, this determines how ``strong'' is the shift of the pixels in the grid. For instance, for a given factor `\(c\)', if \(n\gg c \cdot\langle{\hat{O}_n}\rangle\), then \(n\) is the only factor of the sorting step, and there is no reordering. On the other extreme, if \(n\ll c \cdot\langle{\hat{O}_n}\rangle\), then this is the same as sorting the observables in the first place, which may lead to a more intense and randomized shifting. Having \(n\sim c \cdot \langle{\hat{O}_n}\rangle\) allows the shifting of grid points, while partially preserving the resemblance of the original figure.
To show the reordering process of this painting numerically, it is useful to use a toy example. Consider a $N=4$ spin Ising chain:

\begin{itemize}
    \item At time \(t=0\), we start with the configuration \(\ket{0000}\), and, as seen in Eqs.~(\ref{eq:reorder}-\ref{eq:zero}), this means that \(i_n = n\), thus there is no reordering. 

    \item Now, let us assume that, when $t=\Delta t$, the evolved state is exactly \(\ket{0010}\). We can see in Eqs.~(\ref{eq:example}-\ref{eq:example2}) that the expectation values of \(\hat{O}_n\) and the computed \(i_n\) are:
    
    \begin{equation}
        \label{eq:example}
        \bra{0100}\hat{O}_0\ket{0100} = \bra{0100}\hat{O}_1\ket{0100} = \bra{0100}\hat{O}_3\ket{0100} = 0 \quad ; \quad \bra{0100}\hat{O}_2\ket{0100} = 1,
    \end{equation}

    thus
    
    \begin{equation}
        \label{eq:example2}
        \mathbf{i} = [i_0, i_1, i_2, i_3] = [0, 1, 12, 3].
    \end{equation}
    
    Therefore, the reordered list would be \(\mathbf{\Bar{i}} = [0, 1, 3, 12] = [i_0, i_1, i_3, i_2]\). As a result, the two rightmost pixels are swapped, as summarized in Fig.~\ref{fig:example}.

\end{itemize}

\begin{figure}[htp!]
    \centering
    \includegraphics[width=1\linewidth]{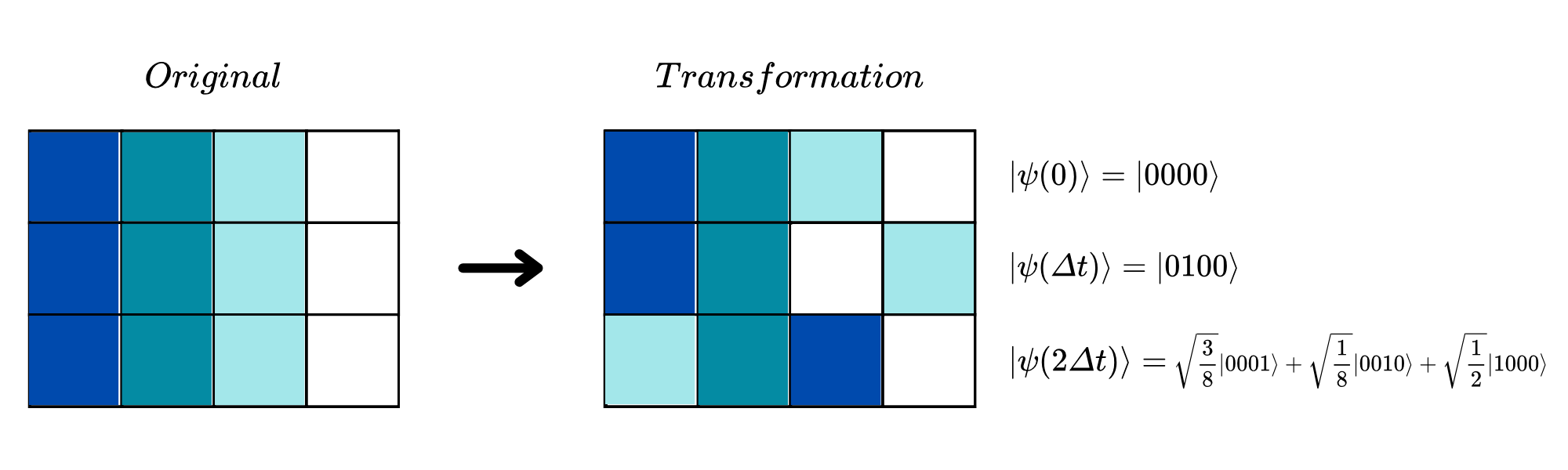}
    \caption{\textbf{Depiction of the transformation of a grid using the time evolution:} Example of a $3\times 4$ grid image being transformed with using Eq.~\eqref{eq:obs_On} and Eq.~\eqref{eq:t2_i}.}
    \label{fig:transform}
\end{figure}

\begin{itemize}
    \item
Finally, assume that on the next time step $t=2\Delta t$, the quantum state is given by the following expression:

\begin{equation}
\label{eq:t2}
    \ket{\psi({2\Delta t})} = \sqrt{\frac{3}{8}}\ket{0001} + \sqrt{\frac{1}{8}}\ket{0010} + \sqrt{\frac{1}{2}}\ket{1000}.
\end{equation}

Then, it follows that the computed list from Eq. (\ref{eq:reorder}) would produce

\begin{equation}
    \mathbf{i} = [\;0\!+\!3.75, \; 1\!+\!1.25, \; 2\!+\!0, \; 3\!+\!5\;] = [\;3.75, \,\;2.25, \;2, \;8\;]
    \label{eq:t2_i}
\end{equation}

and the reordering would be \([i_2, i_1, i_0, i_3]\). As a result, an image grid with 3 rows and 4 columns could be transformed as depicted in Fig.~\ref{fig:transform}.
\end{itemize}

For the quantum computation, we use the \texttt{ibm\_kyoto} quantum hardware with $4096$ shots (measurements of the quantum circuit) and we translate the quantum computing results into new positions of the squares in the grid. Fig.~\ref{fig:narcissus} shows the photograph of the final oil painting, hand-painted on a wooden panel\footnote{This painting has been exhibited at the European Quantum Technologies Conference (EQTC) in Lisbon (Portugal)~\cite{eqtc}.}.

\section{Quantum Transformation II: Magritte}\label{sec:magritte}
In this second masterpiece, painted in 1964 by the Belgian surrealist René Magritte and with the title ``\textit{Le fils de l'homme}''~\cite{magritte}, we have again a portrait of a man, but the reflection here is on the entire painting in itself, as this is a self-portrait of the artist.
The composition consists of a man in an overcoat wearing a bowler hat. He is standing in front of a wall overlooking water and a cloudy sky. A green apple hides the man's face and the man's left arm appears to bend backwards at the elbow. 
Magritte said about the painting that everything we see hides [something else], we always want to see what is hidden by what we see. There is an interest in that which is concealed and which the visible does not show us. This interest can take the form of a quite intense feeling, a sort of conflict, one might say, between the visible that is hidden and the visible that is present~\cite{magritte1977magritte}.

% \begin{figure}[H]
%     \centering
%     \caption{\textbf{The painting ``\textit{Quantum Transformation II: Magritte}'' (oil on wooden panel):} In the oil painting, the entire subject is modified by the quantum time evolution. The only element that remains untouched is the green apple. \textit{Panel size: $60\times 77$ cm.}  }  
%     \label{fig:magritte}
% \end{figure}

Also in this painting, we refer to the concept of observation and subsequent interaction with a quantum system. 
In our quantum aesthetics version\footnote{The original image is subject to copyright restrictions and cannot be reproduced here.}, we apply the quantum process to the entire painting (again, as the conceptual surface of reflection) displacing all of its elements, with the exception of the apple in front of the man's face. With this process of displacement, we are led to imagine the features of the face that now become unveiled. 
To create the man's face, we edit the digital image of the painting by taking inspiration from another painting by Magritte, ``\textit{La bonne foi}''~\cite{labonnefoi}.
To reach this result, we divide the painting into 16 columns and 20 rows. 
We use a similar approach as in Sec.~\ref{sec:narciso}, with the computation of Eq.~\eqref{eq:obs_On} and the subsequent reordering of the elements.
Since the computation can be quite demanding, we decided to mirror the time evolution with 10 steps, so we can reduce the number of Trotter steps and the depth of the circuits. 
In result, the painting's upper half utilizes the regular time evolution (going from the first time-step in the bottom and reaching the middle at step 10), whereas the bottom half is mirrored (the middle section starting from step 10, down to one in the last row). Consequently, the middle section of the painting contained the most time-evolved content. Furthermore, instead of duplicating a single 10-step result for both halves, the final computation counted with two independent runs of the same circuit, both simulating 10 Trotter steps. 
The upper half was computed with \texttt{ibm\_sherbrooke},  and the bottom one with \texttt{ibm\_strasbourg}; both with 4096 shots per circuit.

\section{Quantum Transformation III: Richter}\label{sec:richter}

\begin{figure}[htp!]
    \centering
     \begin{subfigure}[b]{0.49\textwidth}
         \centering
         \includegraphics[width=\textwidth]{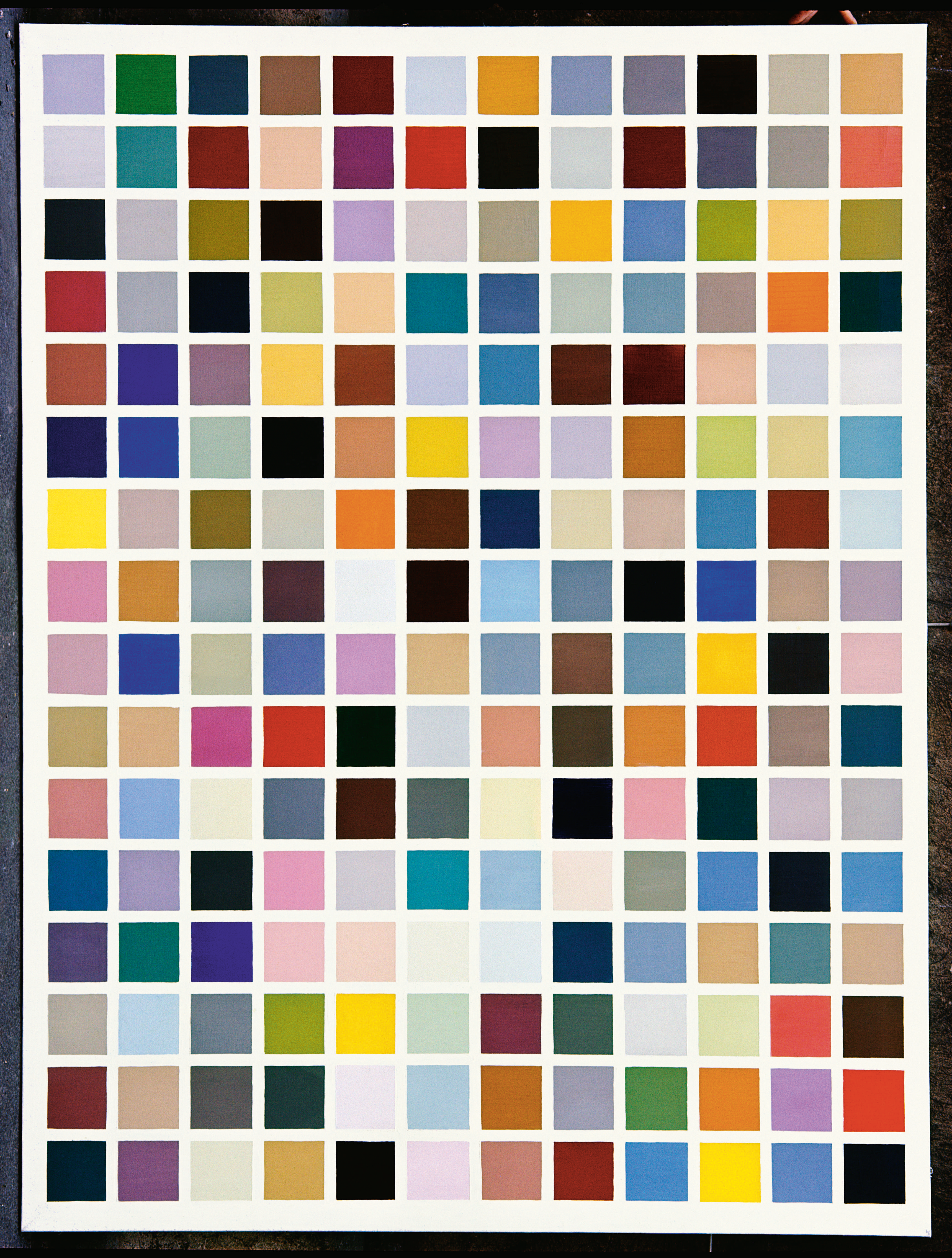}
         \caption{}
         \label{fig:richter_og}
     \end{subfigure}
     \begin{subfigure}[b]{0.48\textwidth}
         \centering
         \includegraphics[width=\textwidth]{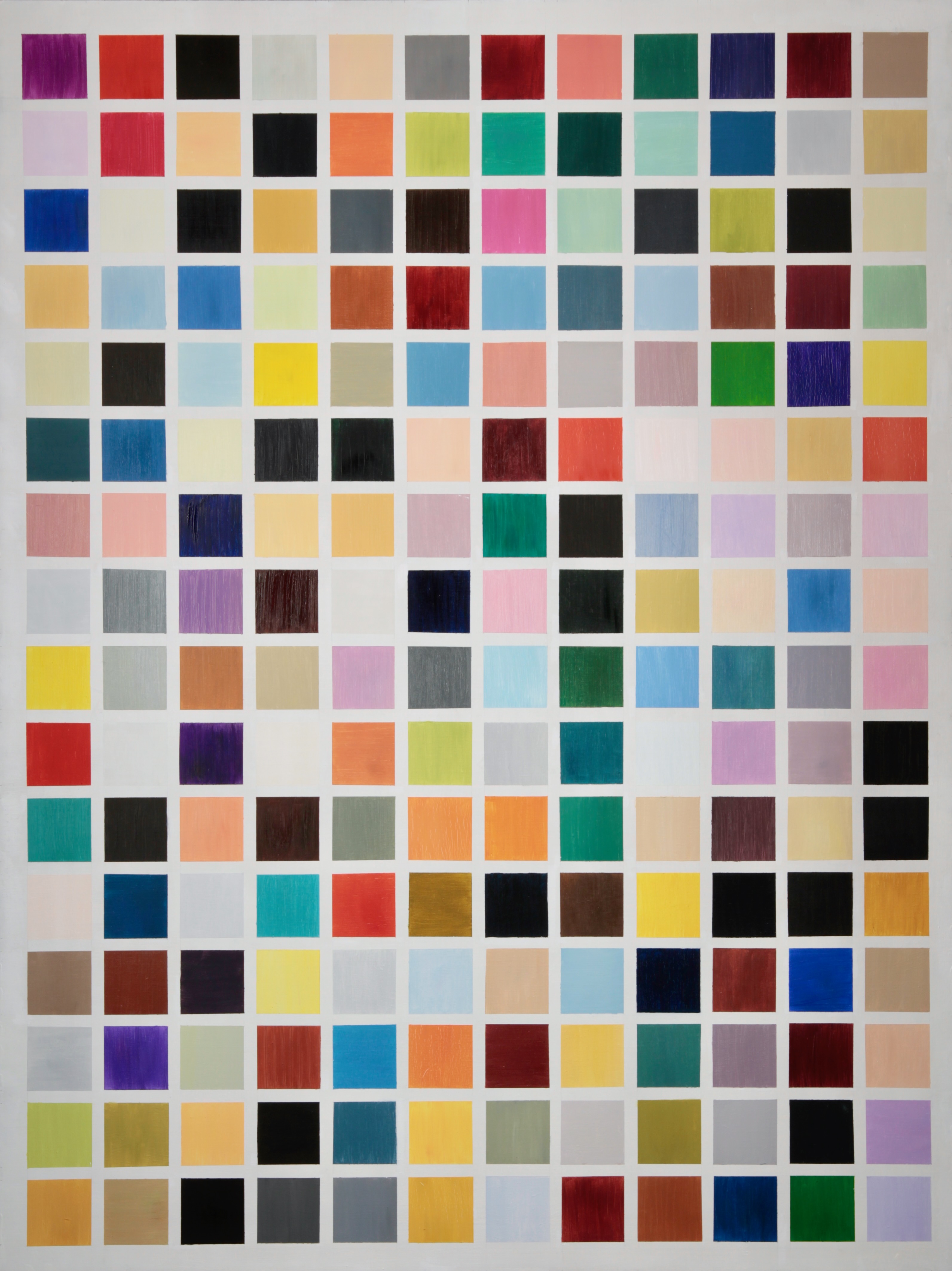}
    \caption{ }
         \label{fig:richter}
     \end{subfigure}
     \caption{\textbf{The painting ``\textit{Quantum Transformation III: Richter}'':} (\textit{panel (a)}) The original version titled ``\textit{192 Farben}'': The image of the original painting. \textcopyright Gerhard Richter 2025 (0085). (\textit{panel (b)}) The revisited version (oil on wooden panel): This photograph of the oil painting illustrates the colors changed by translating the results from quantum computation. \textit{Panel size: $75\times 100$ cm.} }
\end{figure}

In this section, we consider the work of the German painter Gerhard Richter and in particular his piece ``\textit{192 Farben}''~\cite{192colors}, painted in 1966. In this case, there is no recognizable shape, only squares and bright colors. 
This painting belongs to a series called \textit{color Charts}. 
On a visual level, the color Chart paintings
are abstractions; but they are also representations of objects, that is, hypothetical examples of the color sample card. 
Richter began applying the chart procedure of selecting the colors at random, further removing the artist’s engagement in the compositional process. In our project, we completely remove the artist's decision and the ``randomness'' is decided by only the quantum computation.
Richter’s statement is that the color Charts are more appropriately designated as Pop art - a movement in which the artist was a major player on the European art scene~\cite{richter}.

The painting already has a clear lattice structure of 12 columns and 16 rows. Thus, we borrow it for structuring the quantum computing process. 
In this painting we aim to reorder elements across all time slices, rather than just in a single one, and use the Hamiltonian, Eq.~\eqref{eq:ising_H}, with all couplings ($J_n,h_{z,n},h_{x,n}$) equal to $1$. We encode the color index $C_n(t) \in [0, 192)$  using the expectation value of the observable in Eq.~\eqref{eq:obs_On}: 
\begin{equation}
    C_n(t) = \bra{\psi(t)} \hat{O}_n \ket{\psi(t)} \cdot 192.
\end{equation}
Differently from Sec.~\ref{sec:narciso} and Sec.~\ref{sec:magritte}, the index \(n\) is not used in the reordering, as the intention is to produce a more intense rearrangement of the colors across the entire painting (see the explanation in Sec.~\ref{sec:narciso}).
The initial state $\ket{\psi(0)}$ is prepared by single $R_y(\theta_n)$ rotation gates to let $C_n(0)  = n$. After computing the color indices for all the time slices and sites, we reorder these 192 values from smallest to largest. The corresponding colors are then rearranged from left to right and top to bottom, following the new order of $C_n(t)$. As the initial color indices at $t=0$ are relatively small, they remain in the original position, i.e. the bottom row of the figure, and other color squares are reordered based on $C_n(t)$. 
We use the \texttt{ibm\_nazca} quantum hardware for the quantum computation and again 4096 shots.
After the time evolution, with Eq.~\eqref{eq:u_t_approx}, the final resulting artwork depicts the same colors of the original composition (Fig.~\ref{fig:richter_og}) but reordered with quantum results (Fig.~\ref{fig:richter})\footnote{Note that possible imperfections in the real physical painting are due to the challenge of exactly reproducing the colors by hand, as well as the difficulty of exactly reproducing the original colors in RGB.}.

\section{Conclusion and outlook}\label{sec:conclusion}

In this paper we have described new ways to combine quantum computer-generated digital images with the artistic practice of oil painting. In particular, we have revisited masterpieces from Caravaggio, Magritte and Richter, spanning different styles of paintings. By discretizing images of the painting and using a quantum algorithm, running on real quantum hardware, a quantum reordering of the lattice tiles was achieved, leading to quantum transformed images of the original paintings. Those were then taken as templates for a reproduction with oil painting on a wooden panel. The obtained oil paintings constitute thus a result of combining the quantum mechanical world of superposition and entanglement, which escapes our daily experience, and real-world painting with brush and canvas. Thus, we have achieved a confluence of both technical and artistic work that can contribute to the ongoing discussion about Quantum Aesthetics~\cite{QuantumAestheticsISQCMCfocusgroup}.

The idea of combining quantum computers and human intervention can be extended in many different directions. Here, we only used one particular algorithm -- real time evolution through a Trotterization. One could think of many other algorithms, sampling, variational methods, quantum machine learning, quantum differential equation solvers and many more. In addition, as in the case of real-world oil painting, there are many possible subjects for which a qubit encoding can be explored to be used in a particular quantum algorithm. 
Moreover, as mentioned above, the subjects chosen define a goal of this project: build a continuous line from realism to abstraction, through the manipulation with quantum computing. 
We believe that the idea of interfacing quantum computing technology and human interaction can lead to novel ways to look at Art. Conversely, employing quantum algorithms for this purpose can help us to learn more about how the inner mechanism of the quantum mechanical world is operating, thus leading to a better understanding of the principles of superposition and entanglement.

\begin{acknowledgments}
We acknowledge the use of IBM Quantum services for this work. The views expressed are those of the authors, and do not reflect the official policy or position of IBM or the IBM Quantum team.
We are grateful to Stefan K\"uhn for his help in photographing the paintings.
A.C.\ is supported in part by the Helmholtz Association — “Innopool Project Variational Quantum Computer Simulations (VQCS).” O.C.H. is supported by the European Project IIMPAQCT, under grant agreement Nr. 101109258 (DOI \texttt{10.3030/101109258}).
This work is supported with funds from the Ministry of Science, Research and Culture of the State of Brandenburg within the Centre for Quantum Technologies and Applications (CQTA). 
\begin{center}
    \includegraphics[width = 0.1\textwidth]{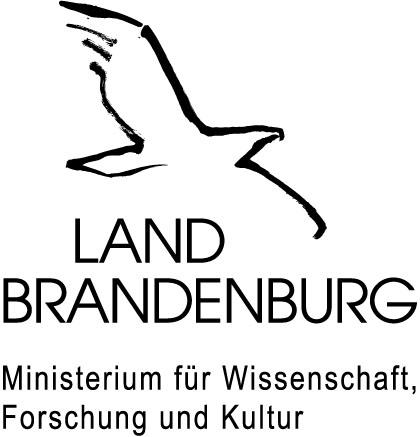}
\end{center}
This work is funded by the European Union’s Horizon Europe Frame-work Programme (HORIZON) under the ERA Chair scheme with grant agreement no. 101087126.
\end{acknowledgments}

\bibliography{bibliography}

\end{document}